\documentclass[twocolumn,prb,superscriptaddress,showpacs]{revtex4}

\usepackage{graphics}
\usepackage{graphicx}
\usepackage{color}
\usepackage{amssymb,amsmath}
\usepackage{multirow,bm}
\usepackage{subfigure}

\newcommand{\e}{\varepsilon}

\newcommand{\x}{\xi}
\renewcommand{\d}{\delta}

\newcommand{\w}{\omega}

\renewcommand{\Im}[1]{\mathrm{Im}\left(#1\right)}

\newcommand{\vk}{\mathbf{k}}

\newcommand{\vR}{\mathbf{r}}

\newcommand{\D}{\Delta}

\newcommand{\hg}{\widehat{g}}

\newcommand{\hG}{\widehat{G}}

\newcommand{\hT}{\widehat{T}}
\newcommand{\hA}{\widehat{A}}
\newcommand{\CG}{\check{G}}

\newcommand{\CU}{\check{U}}
\newcommand{\bU}{\bar{U}}
\newcommand{\CT}{\check{T}}

\renewcommand{\t}{\widehat{\tau}}

\renewcommand{\(}{\left(}
\renewcommand{\)}{\right)}

\def\DHLhksqrt#1#2{%
\setbox0=\hbox{$#1\sqrt{#2\,}$}\dimen0=\ht0
\advance\dimen0-0.2\ht0
\setbox2=\hbox{\vrule height\ht0 depth -.75\dimen0 width .2\ht0}%
{\box0\lower0.4pt\box2}}

\begin{document}

\title{Impurity states in multiband $s$-wave superconductors: analysis of iron pnictides}



\author{R. Beaird}
\affiliation{Department of Physics and Astronomy, Louisiana State
University, Baton Rouge, LA 70803-4001, USA}
\author{I. Vekhter}
\affiliation{Department of Physics and Astronomy, Louisiana State
University, Baton Rouge, LA 70803-4001, USA}
\author{Jian-Xin Zhu}
\affiliation{Theoretical Division and Center for Integrated Nanotechnologies, Los Alamos National Laboratory,
Los Alamos, NM 87545, USA}

\date{\today}

\begin{abstract}
We examine the effect of a single, non-magnetic impurity in a multiband, extended $s$-wave superconductor allowing for anisotropy of the gaps on the Fermi surfaces. We derive analytic expressions for the Green's functions in the continuum and analyse the conditions for the existence of sharp impurity-induced resonant states. Underlying band structure is more relevant for the multiband than for single band case, and mismatch between the bands generically makes the formation of the impurity states less likely in the physical regime of parameters. We confirm these conclusions by numerically solving the impurity problem in a tight-binding parameterization of the bands relevant to pnictide superconductors.
\end{abstract}

\maketitle

Since the discovery of the multiband iron-based superconductors (FeSC)~\cite{KamiharaSCinLaFePO:2006,KamiharaSCinLaFeAsO:2008,RotterSCinBaFe2As2:2008,ChenSmFeAsOF:2008,Hsu:2008,SubediDftFeX:2008}, the determination of their gap shape has been an 
active subject of study. Several theories proposed that the  superconducting order parameter for many pnictides and dichalcogenides is of the $s$-wave ($A_{1g}$) type.
However the presence of several Fermi surface (FS) sheets away from the center ($\Gamma$-point) of the Brillouin Zone (BZ) allows for strongly anisotropic, or nodal (with zeroes) 
gap.  Experiments sensitive to low energy quasiparticle excitations indicate that the gap structure is non-universal across the families and doping ranges, so that complementary methods have to be employed for a complete picture to emerge.

These measurements include the temperature dependence of the penetration depth and spin-lattice relaxation rate, both temperature and field variation of the specific heat and the thermal conductivity, and angle-resolved photoemission, among others. Scanning tunneling spectroscopy (STS) at or near impurity sites 
played an important role in testing the anisotropic superconducting state of the high-T$_c$ cuprates~\cite{BalatskyImpRMP:2006}, and therefore it is natural to ask what information it can provide for the iron-based materials. In conventional single-band superconductors with an isotropic gap, potential scatterers do not change the local density of states appreciably. In contrast, for superconductors with sign-changing order parameter on the Fermi surface, non-magnetic impurities create a quasi-bound (resonant) state, whose energy relative to the gap maximum and the shape in real space both carry information about the gap shape.

Existence of multiple FS sheets complicates the picture, and several recent studies arrived at different conclusions. For tight-binding bands with the extended $s$-wave $\cos(k_x)\cos(k_y)$  gap (in the ``unfolded'' BZ, see below), Refs.~\onlinecite{TsaiImpurity:2009, ZhouImpuritySx2y2BdG:2009} do not find low-energy impurity states, and Ref.~\onlinecite{ZhangNonmagSignReverse:2009} finds such states
at or above the energies of about half of the SC gap amplitude. In two-band continuum models of isotropic gaps of opposite signs ($s_{+-}$), Refs.~\onlinecite{MatsumotoSingleImpSC:2009, NgNonMagImpSMS:2009} find that the impurity resonances can form deep within the gap.
The Bogoliubov-de Gennes analysis of the five-band model ~\cite{KariyadoSingImpSpm:2010} finds the low energy impurity states for some intermediate values of the scattering potential. The inevitable question is how general the results obtained using a specific set of assumptions are, and whether different results are due to details of the models or salient features of  pairing.

We address this issue in the current Communication. We combine analytical and numerical techniques to investigate the impurity resonance states
for different gap shapes in the $A_{1g}$ representation. One of our conclusions is that, for similar gap shapes, details of the band structure, even for similar Fermi surface topologies, strongly affect the location of the resonance state. Also, the role of inter- and intra-band impurity scattering potentials in the formation of the impurity resonance is different depending on the degree of anisotropy in the gap. We explain the physics behind these effects by combining analytical and numerical approaches.

The simplest model of FeSC superconductors that captures the relevant physics has two FS sheets: one hole-like (h) around $\Gamma$-point and one electron-like (e) close to $M$ and equivalent points in the BZ. It allows the analysis of the two principal scenarios for the $A_{1g}$ pairing: a) $s_\pm$ where magnetically-assisted predominantly interband pair scattering requires  a sign change in the order parameter between FS sheets of different type~\cite{MazinUnconSpmSC:2008}, and b) $s_{++}$  where orbital fluctuations promote pairing with the same sign on the FS sheets~\cite{KontaiOrbFluctSpp:2010, SaitoOrbFlucts:2010, YanagiSppSpmMagFlucts:2010}. Under both pictures the gap on the hole sheet is nearly isotropic in the x-y plane, while Coulomb repulsion may lead to an anisotropic, or even nodal, gap on the electron sheet. %

To describe scattering by a non-magnetic impurity we make a simplifying assumption that the scattering amplitude depends only on the band index of the initial and final states, so that $H_{imp}=\sum_{\vk \vk'\sigma}U_{jj'} c_{j\vk\sigma}^{\dagger}c_{j'\vk'\sigma}$, where $U_{jj'}=U_0$ if the band indices $j=j'$ ($j=e,h$) and $U_{jj'}=U_1$ otherwise. The
approximate independence of the elements of the $4\times 4$-matrix $\check{U}$ (we use $\widehat{U}$ to denote $2\times 2$ matrices in Nambu space) is justified by the small size of the FS in FeSC. This parametrization means separation into small and large momentum transfer scattering, and naively one can expect $U_1\ll U_0$ because of screening. However band structure calculations show that the same Fe $d$-orbitals contribute significantly to both the electron and the hole sheets of the Fermi surface, ~\cite{GraserPairingDegeneracy:2009} and hence an impurity at or near the Fe site, will produce a significant inter-band scattering component. One candidate is the Co-dopants in the 122 series~\cite{KemperCoDoping:2009}.

We compute the Green's function which is a matrix in both band and particle-hole space,
$\CG_{jj^\prime}(\vk,\vk^\prime; \tau) =
  -\left<T_\tau\left[\Psi_{j\vk}(\tau)\otimes\Psi^\dagger_{j^\prime\vk^\prime}(0)\right]\right>\,,
$
where the Nambu spinor is $\Psi_{j\vk}^\dagger$=$(c_{j\vk\uparrow}^\dagger,c_{j-\vk\downarrow})$, and $T_\tau$ is the imaginary time ordering operator. In this notation the Hamiltonian for a pure superconductor is $\widehat{H}_{j\vk}$=$\xi_{j\vk}\t_3+\D_{j\vk}\t_1$, with $\t_0$ the identity matrix and $\t_i$ ($i$=1$\ldots$3) the Pauli matrices in the Nambu space, $\xi_{j \vk}$ is the quasiparticle energy in band $j$, and $\D_{j\vk}$ is the superconducting gap function on the $j$-th Fermi surface sheet. We ignore the weak dispersion along the $c$-axis.
In the absence of impurities
$\CG_{0,jj^\prime}(\vk,\vk^\prime)$=$\delta_{jj^\prime}\delta_{\vk\vk^\prime}\hG_{0,j}(\vk)$, with 
$\hG_{0,j}(\vk;i\w_n)$=$(i\w_n\t_0-\widehat{H}_{j\vk})^{-1}$, 
%
and Matsubara frequencies are $\w_n=2\pi T(n+1/2)$.

For an extended $s$-wave $\D_{\vk}=A+B\left[\cos(k_xa)+\cos(k_ya)\right]$, in the ``unfolded'' zone scheme, with $a$ is the lattice constant for the square Fe lattice.~\cite{MazinUnconSpmSC:2008, MazinUnconvSymmReview:2009, GraserPairingDegeneracy:2009,ChubukovKDependAndNodes:2009}. When projected on the hole, $S_h$, and electron, $S_e$, FS sheets this results in a nearly isotropic $\Delta_{h\vk}\approx \Delta_h$ for $\vk\in S_h$, and a generally anisotropic gap on the electron sheet(s),
$\Delta_{e\vk}$=$-\Delta_e (1+r\cos 2\phi)$. Here $\phi$ is the angle as measured from the [100] and [010] directions at $(\pm\pi,0)$ and $(0,\pm\pi)$, respectively.  In our notations $\delta_0=\Delta_e/\Delta_h>0$ for $s_\pm$ state and $\delta_0<0$ for the $s_{++}$ state.
Below we compare the values $r=0$ (isotropic gap on electron Fermi surface), $r=0.7$ (nodeless anisotropic gap), and $r=1.3$ (nodal gap). 

We solve the single impurity problem using the $T$-matrix ($\CT$) method, where $\CT$ and $\CG$ satisfy the coupled equations~\cite{BalatskyImpRMP:2006}
$\CG=\CG_0+\CG_0\CT\CG_0$ and $\CT=\CU+\CU\sum_{\bm k}\CG_0\CT$. For the momentum-independent $\CU$, the solution
$\CT=\left[\check{1}-\CU\sum_{\bm k}\CG_0\right]^{-1}\CU$ is solely a function of the band index and the frequency. Upon analytic continuation $i\omega_n\rightarrow \omega+i0^+$ the poles of $\CG$ give the energies of elementary excitations, and hence the poles of $\CT(\omega)$ give the energies of the impurity-induced states. The density of states per spin in each band is
$N_j(\vR,\w)=-{\pi}^{-1}\Im{\check{G}_{jj,11}(\vR,\vR;\w+i\,0^+)}$, where indices
$11$ refers to the particle component in the Nambu space.

Denoting $\hg_j=\sum_\vk\hG_{0, j}(\vk)=\sum_i g_{ji}\t_i$, with $i=0,\ldots,3$, we find that the components of the $T$-matrix satisfy $\hT_{ee} = U_0\t_3+U_0\t_3\hg_{e}\hT_{ee}+U_1\t_3\hg_h\hT_{he}$ and
$\hT_{he} = U_1\t_3+U_1\t_3\hg_{e}\hT_{ee}+U_0\t_3\hg_h\hT_{he}$; the equations for $T_{hh}$ and $T_{eh}$ are obtained by switching indices. The solution is 
$\hT_{ee}=\hA^{-1}\left[U_0\t_3-(U_0^2-U_1^2)\t_3\hg_h\t_3\right]$
where
$\hA =\(1-U_0\t_3\hg_h\)\(1-U_0\t_3\hg_e\)-U_1^2\t_3\hg_h\t_3\hg_e$. Consequently, the energies of the bound state are determined from $\det(\hA)\equiv D(\w)=0$, where
\begin{equation}\label{D full g3s}
\begin{split}
D(\w)
&=U_0^4 \(g_{e0}^2 - g_{e1}^2-\pi^2 N_e^2c_e^2 - \overline{g_{e}}^2\frac{ U_1^2}{U_0^2}\)\\
 &\hspace{5mm}\times\(g_{h0}^2 - g_{h1}^2-\pi^2 N_h^2c_h^2 - \overline{g_{h}}^2\frac{ U_1^2}{U_0^2}\)\\
 &-U_1^2\(\overline{g_e}^2 + \overline{g_h}^2 + 2(g_{e0}g_{h0} - g_{e1}g_{h1} + g_{e3}g_{h3})\)\,.
\end{split}
\end{equation}
Here we  introduced $\overline{g_{j}}^2$=$\sum_i(-1)^ig_{ji}^2$
and $c_j^2$=$(\pi N_j U_0)^{-2}\(1 - U_0g_{j3}\)^2$. Note that Eq.~\eqref{D full g3s} depends only on 
$U_1^2$, and hence we take $U_1>0$ without loss of generality.

As in single-band superconductors, the gap shape affects the angular averages of the anomalous Green's functions, $g_{j1}$, and therefore influences the energy of the impurity state both in a single- and multi-band cases. Two other aspects reflect the multiband nature of the system. First, since summation over momenta in the $T$-matrix equations yields a prefactor of the density of states (DOS) at the appropriate FS sheet, the ratio of the DOS 
$n=N_e/N_h$ controls the number of states available for inter- vs. intra-band scattering. In the models with dominant interband pairing the same parameter controls the ratio of the gap amplitudes, $\delta_0$.
In analogy with Refs.~\onlinecite{BangPossiblePairingStates:2008,DolgovBCSvsEliashberg:2008} we find that at $T=0$ such models obey the constraint
$n \d_0^2 (1+r^2/2)=1$. 
Variation of $n$ very significantly affects the properties of the impurity bound state.

Second, the particle-hole band asymmetry,
\begin{equation}
  g_{j3}=\frac{1}{2}\sum_{\bm k}\mbox{Tr}\left[\CG_{0,j}\widehat{\tau_3}\right]=-\sum_{\bm k}\frac{\xi_{j\bm k}}{\w_n^2+\x_{j\vk}^2+\D_{j\vk}^2}\,,
  \label{eq:g3}
\end{equation}
appears in combination with the the impurity potential in the equation for the $T$-matrix. This term is largely determined by the normal state band structure, and therefore can be approximated by its value with $\D_{j\vk}=0$. In analytical approaches it is often assumed $g_{j3}=0$, although even in single-band superconductors 
the resonance state is sensitive to the band structure~\cite{FehrenbacherNonMagImpVHS:1996, FehrenbacherNonMagDwaveArpes:1996}. In FeSC the situation is even more complex since the chemical potential is close to the top/bottom of the hole/electron bands respectively. As a result: a) $g_{e3}$ and $g_{h3}$ have opposite signs and hence the sign of $U_0$ 
matters; b) we expect $g_{e3}/N_e\sim - g_{h3}/N_h \sim O(1)$, which changes the bound state properties relative to the particle-hole symmetric case.

\begin{figure}[t]
\includegraphics[width=0.45\textwidth,clip=true]{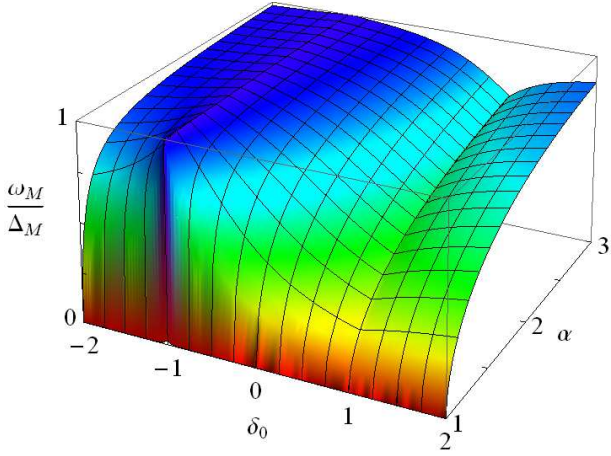}
\vspace{-4mm}\caption{(Color online) Energy of the impurity bound state for isotropic gaps ($r=0$). Here $\w_{\rm M}=\min(\omega_+,\omega_-)$, see Eq.~\eqref{eq:wpm2}, and minimal gap $\D_{\rm M}={\rm min}(\D_e,\D_h)$. For a given value of the DOS ratio $N_e/N_h$, and the corresponding fixed $\delta_0$, variation of the impurity potentials $U_0$ and $U_1$ traces a path along the surface. For a given value of $\alpha$, the bound state energy is always lowest at $\d_0$=1.}
\label{fig:boundS}
\end{figure}

\begin{figure}[t]
\includegraphics[width=0.45\textwidth,clip=true]{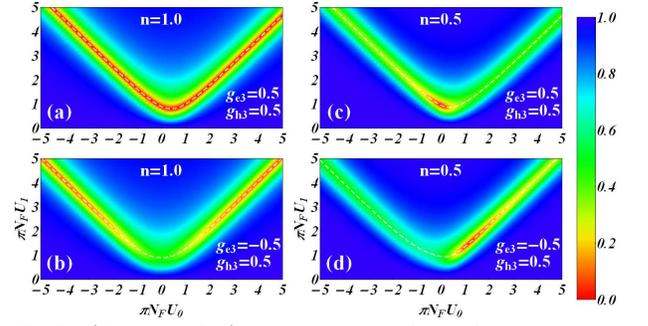}
\vspace{-4mm}\caption{(Color online) Energy of the impurity bound state for $r=0$ for different DOS ratios, $n$, and the particle-hole anisotropies, $g_{j3}$, measured in units of $\pi N_j$. We denoted $N_F=(N_e+N_h)/2$. Note that the deep and well-defined bound state only appears for the strong interband scattering $U_1$.}
\label{fig:boundSpm}
\end{figure}

The location of the bound state for the isotropic gap, $r=0$, is given by 
\begin{equation}\label{eq:wpm2}
  \begin{split}
  \w_\pm^2=&\frac{\alpha^2(\Delta_e+\Delta_h)^2-2(\alpha^2-1)\Delta_e\Delta_h}{2(\alpha^2-1)}
  \\
  &\hspace{2mm}\pm\frac{\alpha(\Delta_e+\Delta_h)}{2(\alpha^2-1)}
\sqrt{\alpha^2(\Delta_e-\Delta_h)^2+4\Delta_e\Delta_h}\,,
  \end{split}
\end{equation}
where all the information about the DOS and scattering potentials is contained in the parameter (note $\alpha\geq 1$)
\begin{equation}\label{alpha with g3s}
\begin{split}
\alpha
=&\frac{\pi^2N_e N_h U_0^4}
{2U_1^2}\left[1+c_e^2-\(1+\frac{g_{e3}^2}{\pi^2N_e^2}\)\frac{ U_1^2}{U_0^2}\right] \\
&\hspace{1.5cm}\times\left[1+c_h^2-\(1+\frac{g_{h3}^2}{\pi^2N_h^2}\)\frac{ U_1^2}{U_0^2}\right]\\
&+\frac{\pi^2(N_e^2+N_h^2)+(g_{e3}-g_{h3})^2}{2\pi^2N_e N_h}\,.
\end{split}
\end{equation}
Fig.~\ref{fig:boundS} shows that a) the bound state is much deeper in the gap for the $s_{+-}$ than for the $s_{++}$ state; b) the lowest energy of the bound state is reached for the ``balanced'' band case, $N_e=N_h$ and $\Delta_h=\Delta_e$, and any deviation from this regime leads to the impurity state edging closer to the continuum. Fig.~\ref{fig:boundSpm} shows several traces over this surface for fixed $n$, where the resonance state becomes closer to the gap edge and exists in a narrower and narrower range of parameters as the densities of states in the two bands ``detune''. Hence not only the topology of the Fermi surfaces, but also the  curvature of the bands at the Fermi level and the relative bandwidth matter. In our view it is for that reason that the simplest tight-binding parameterization~\cite{RaghuMinimal2bandModel:2008} does not give a clearly defined bound state. For that model we estimate $N_e(0)\approx4.9N_h(0)$,  $g_{e3}\approx-3.7\pi N_e(0)$, and  $g_{h3}\approx0.4\pi N_h(0)$, which suggests that the impurity state is very close to the gap edge, or does not form at all, in agreement with the conclusions of Refs.~\onlinecite{TsaiImpurity:2009,ZhouImpuritySx2y2BdG:2009}. In the extreme case of the proximity to van Hove singularity for one of the bands, the impurity states form only for $U_1\gg U_0$ ~\cite{Karol}. In contrast, the studies utilizing more realistic band structure fits with smaller ratio of the DOS often find the deeper and sharper impurity resonance~\cite{NgNonMagImpSMS:2009}.

For $r=0$ the lowest energy of the impurity resonance is reached for
\begin{equation}\label{U1m spm g3}
\frac{U_{1,m}^2}{U_0^2}=
\sqrt{\frac{1+c_e^2}{1+\(\frac{ g_{e3}}{\pi N_e}\)^2}}
\sqrt{\frac{1+c_h^2}{1+\(\frac{ g_{h3}}{\pi N_h}\)^2}}\,.
\end{equation}
For particle-hole symmetric case ($g_{e3}=g_{h3}=0$) this clearly indicates $U_1>U_0$, which is unlikely. For strongly particle-hole asymmetric bands, and for strong scattering ($c_{e,h}\ll 1$), the impurity bound state may be formed for more physical values of $U_1\leq U_0$, but, as is seen from Fig.~\ref{fig:boundS}, the state itself is at finite energy.
%

\begin{figure}[t]
\includegraphics[width=0.45\textwidth,clip=true]{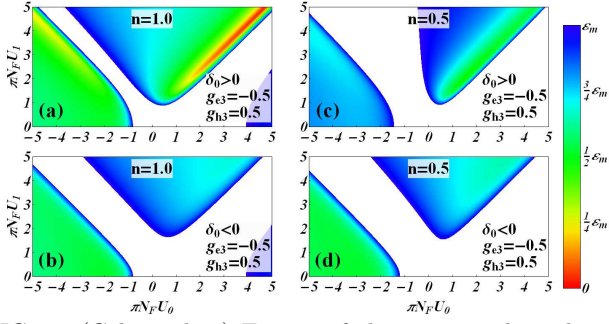}
\vspace{-4mm}\caption{(Color online) Energy of the impurity bound state for $r=1.3$ for different DOS ratios, $n$, and the particle-hole anisotropies, $g_{j3}$, in the same notations as in previous figure.}
\label{fig:bounddpm}
\end{figure}

As the value of $r$, and the anisotropy of the gap on the electron sheet increases this trend remains for as long as the gap does not develop nodes. Once the gap on the electron sheet develops zeroes, a more familiar aspect of the single-band nodal superconductors comes into play: intraband scattering produces a bound state, while the interband scattering broadens it. This is illustrated in Fig.~\ref{fig:bounddpm}. Even in this case the small size of the Fermi surfaces, i.e. moderately large and opposite in sign values of $g_{j3}$ lead to a much shallower resonance state that for a particle-hole symmetric case. Note also a very significant anisotropy with respect to the sign of the intraband scattering potential. These analytical results show that in multiband systems with opposite nature of the carriers in the two bands it is generically difficult to realize a well-defined impurity resonance states.

\begin{figure}[t]
\includegraphics[width=0.46\textwidth,clip=true]{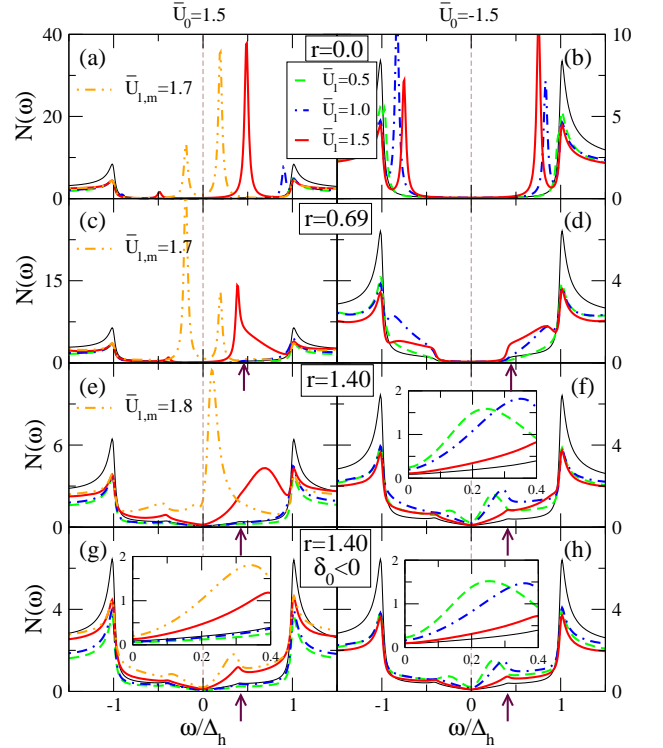}
\vspace{-4mm} \caption{(Color online)  Evolution of on-site LDOS (eV$^{-1}$) for four-band model with moderate scattering, $\bU_0$=$\pm$1.5, for (a,b) isotropic, (c,d) nodeless anisotropic, and (e,f) nodal gaps for $\d_0>0$. Panels (g,h) are for nodal gap with $\d_0<0$. The thin black line is the DOS of clean system, and arrows mark the DOS feature at $\e_m=(0.38,0.41)$ for the anisotropic cases with $r=(0.7,1.4)$, respectively. Low-energy impurity states form below $\e_m$ in the nodal system even at small values of $\bU_1$, but impurity states do not form near $\w$=0 in either of the fully-gapped systems. LDOS for $\bU_{1,m}$ is shown when $\bU_{1,m}\not\approx|\bU_0|$ but is unlikely when $\bU_{1,m}>|\bU_0|$. The insets show close ups of the low-intensity positive-bias peaks.   Note the different vertical scales.}
\label{fig:Korsh Mod}
\end{figure}

To verify that these conclusions remain valid for pnictides we numerically solved the $T$-matrix equation and determined the local density of states on the impurity site for a tight-binding fit to the Fermi surface from Ref.~\onlinecite{KorshunovMagExcite4Band:2008}. In the folded (2-Fe) BZ  the energies of the hole and the electron bands are given by $\xi_{\alpha_i\vk}=-t_{\alpha_i}\(\cos\(k_x\)+\cos\(k_y\)\)
-t^\prime_{\alpha_i}\cos\(k_x\)\cos\(k_y\)-\mu_{\alpha_i}$, and $\xi_{\beta_i\vk}=-t_{\beta i}\(\cos\(k_x\)+\cos\(k_y\)\)
-t^\prime_{\beta_i}\cos\(\frac{k_x}{2}\) \cos\(\frac{k_y}{2}\)-\mu_{\beta_i}$ respectively, with the hoppings and the band shifts (in eV) $(t_{\alpha_1},t^\prime_{\alpha_1},\mu_{\alpha_1})=(-0.3,-0.24,0.6)$ and $(t_{\alpha_2},t^\prime_{\alpha_2},\mu_{\alpha_2})=(-0.2,-0.24,0.4)$ for the hole, and
$(t_{\beta_1},t^\prime_{\beta_1},\mu_{\beta_1})=(-1.14,-0.74,-1.70)$ and $(t_{\beta_2},t^\prime_{\beta_2},\mu_{\beta_2})=(-1.14,0.64,-1.70)$ for the electron bands. In this model $n\approx0.4$, close to the $n=0.5$ considered above, and different from $n\approx 0.2$ used in Refs.~\onlinecite{TsaiImpurity:2009,ZhouImpuritySx2y2BdG:2009}, even though the Fermi surface topology is similar. Hence the difference between the results stems from this DOS imbalance.  Our Green's functions $g_{e3}\approx-\pi N_e(0)$, and  $g_{h3}\approx 1.2\pi N_h(0)$. We again consider the isotropic gap $\D_{\alpha_i\vk}$=$\Delta_h$ on the hole FS sheets, and we define the gap on the electron sheets to be
$\D_{\beta_{1,2}\vk}=-\D_e\(1\pm\tilde{r}\cos\(\frac{k_x}{2}\)\cos\(\frac{k_y}{2}\)\)$, where the upper and lower signs arise from the folding of the Fermi surfaces into the smaller 2-Fe Brillouin Zone. In general the gap anisotropy on the electron FS sheets is not a simple cosine, but is close to it in shape. We can extract the effective anisotropy from the ratio $|\Delta_{e,min}|/|\Delta_{e,max}|\equiv |r-1|/|r+1|$, and make comparison with our analytical results. The electron Fermi surfaces in this description are very close to each other, so that the values of $r$ differ by about 6\% between them, and we quote the average number.
The calculations were performed on a 2000$\times$2000 $k$-space grid with the intrinsic broadening $\gamma$=$\D_h/40$=0.0015 eV.
Fig.~\ref{fig:Korsh Mod} shows the LDOS at the impurity site for moderate values of the scattering potentials. As before, the particle-hole band asymmetry causes the impurity-state energy to depend strongly on the sign of $U_0$.  For these moderate values of $U_0$ and $U_1$ the low-energy states do not form when the FS sheets are fully-gapped ($r=0$) except for the unphysical case $U_1\geq U_0$.  The impurity resonance forms more easily when there are nodes on the electron sheets; this happens regardless of the sign of $\d_0$ since the formation of the state is dominated by intraband processes. For anisotropic nodeless gap ($r=0.7$) there exist broad features associated with the transfer of spectral weight from the coherence peak to energies above the threshold $\varepsilon_m=\Delta_e|r-1|$. These states  mix with the continuum and are not sharp. At the same time we find that they can feature relatively prominently at the impurity site simply because the peak at $\varepsilon_m$ is much smaller than that at $\omega=\Delta_h$ in the pure system. A sharp bound state close to mid-gap only appears for extremely high values of $U_0\simeq U_1\sim 100$, also supporting the qualitative analysis above.

Our main conclusions therefore are that within a model of relatively momentum-independent inter- and intra-band scattering in multiband systems a) the mismatch of the densities of states and the gap values on different Fermi surface sheets affects very significantly the energy of the impurity bound states, pinning them relatively close to the continuum states over most of the parameter range; b) this conclusion remains qualitatively valid even for nodal gaps on one of the Fermi surface sheets for realistic particle-hole anisotropies in the electron and hole bands; c) broad features due to impurities may exist for anisotropic gaps in the $s_{\pm}$ case, but are far less likely for the $s_{++}$ pairing. We explain the differences in the results between different groups as stemming in part from the different underlying band structure. Of course, in a fully microscopic theories starting from the  orbital representation the resulting effective scattering may be anisotropic within each Fermi surface sheet: strong between parts with similar orbital content and weak between regions stemming from different Fe orbitals. It would therefore be very instructive to check such effective potential, for example, in the model of Ref.\onlinecite{KariyadoSingImpSpm:2010}. It will also be useful to check whether within these models the scattering potential varies strongly on the scale of the bandwidth, since such an effect assists the formation of the bound state even for an isotropic gap. Our results clearly show that the study of impurity states alone is not sufficient to draw reliable conclusions about the shape of the superconducting gap in multiband systems.

We are grateful to P. J. Hirschfeld and H.~Takagi for helpful discussions.
Work at LANL was performed under the auspices of the
U.S.\ DOE contract No.~DE-AC52-06NA25396 through the LDRD program, and in part by the Center for Integrated Nanotechnologies, a U.S. DOE Office of Basic Energy Sciences user facility.
Work at LSU was supported by DOE Grant DE-FG02-08ER46492 (R. B. and I. V.).

\end{document}